# Accelerated rogue waves generated by soliton fusion at the advanced stage of supercontinuum formation in photonic crystal fibers.

Rodislav Driben,[1,*] and Ihar Babushkin[2]

[1] Department of Physical Electronics, School of Electrical Engineering, Faculty of Engineering,
Tel Aviv University, Tel Aviv 69978, Israel
[2] Weierstrass Institute for Applied Analysis and Stochastics Mohrenstr. 39, 10117, Berlin, Germany
*Corresponding author: driben@post.tau.ac.il

Soliton fusion is a fascinating and delicate phenomenon that manifests itself in optical fibers in case of interaction between co-propagating solitons with small temporal and wavelengths separation. We show that the mechanism of acceleration of trailing soliton by dispersive waves radiated from the preceding one provides necessary conditions for soliton fusion at the advanced stage of supercontinuum generation in photonic crystal fibers. As a result of fusion large intensity robust light structures arise and propagate over significant distances. In presence of small random noise the delicate condition for the effective fusion between solitons can easily be broken, making the fusion induced giant waves a rare statistical event. Thus oblong-shaped giant accelerated waves become excellent candidates for optical rogue waves.

Since the report of rogue wave analog in optics [1], a substantial progress has been made in understanding of mechanisms behind this dramatic phenomenon [2 - 9]. There are two major competing families of explanation. The first, supercontinuum generation (SC) [10 - 12] family offers strongest Raman-shifted [13] solitons or alternatively products of soliton collisions [4 - 6] as possible candidates for rogue waves. A single strong red-shifted soliton resulting from fission is a common event in SC. Exact dynamics of solitons collisions is hard to predict, providing additional degree of uncertainty. The peak amplitude during collisions significantly exceeds that of contributing single solitons, but such a peak is very short lived. The second family of explanations [7, 8] is based on Akmediev breathers [14] and in particular on a single-peaked solution called the Peregrine's soliton [9]. These works provide waves arising from instability modulation and disappearing without a trace, which is very consistent with the behavior of famous ship killers.

The objective of this work is to present a soliton fusion, occurring at the advanced stage of SC formation in PCF. A large robust light structure resulting from this fusion becomes a very perspective candidate for the SC family of optical rogue waves that possesses strong amplitude dominance over the background for significant propagation distance and also has an oblong, "water wall"-like shape (Fig.1). Moreover the reported effect requires delicate matching of the colliding solitons parameters and therefore it is indeed rarely observed during the SC generation.

A generalized nonlinear Schroedinger equation (GNLSE) including Raman and shock terms [10, 15] was used to model the pulse propagation for full mathematical description see [16, 17]). We consider propagation of ultrafast pulses dynamics in a typical PCF with zero dispersion wavelength $\lambda_0$ = 790 nm. At the wavelength of 800 nm, where the input pulse is centered, the nonlinear coefficient is $\gamma$ = 80 W$^{-1}$km$^{-1}$, and the dispersion coefficients up to seventh order are: $\beta_2$ = -2.1 fs$^2$/mm, $\beta_3$ = 69.83 fs$^3$/mm, $\beta_4$ = -73.25 fs$^4$/mm, $\beta_5$ = 191.95 fs$^5$/mm, $\beta_6$ = -727.13 fs$^6$/mm, $\beta_7$ = 1549.4 fs$^7$/mm.

First, a simple fusion process [18] between two fundamental solitons of sech shape, co-propagating in the PCF with small initial temporal separation of 0.3ps was simulated by solving numerically the GNLSE equation. Fig.1 clearly demonstrates an oblong wave crest, maintaining for several meters of propagation distance, arising due to long interaction between two pulses. The inset displays an image of oceanic rogue wave with it's typical prolong wave crest shape, taken from www.naviguesser.com.

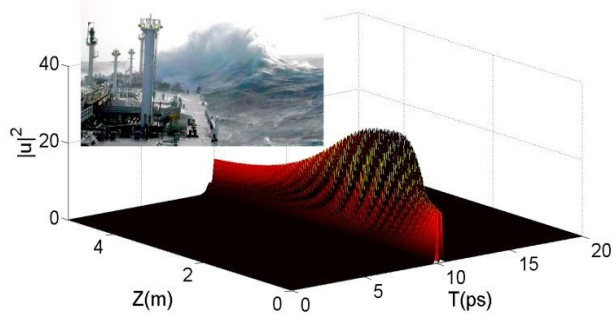

Fig. 1. (Color online) The fusion resulting from co-propagation of two fundamental solitons with wavelengths $\lambda_1$ = 820 nm and $\lambda_2$ = 755 nm. Initial pulses separation is 0.3 ps and their widths $T_{FWHM}$ = 105 fs. The inset displays image of oceanic rogue wave demonstrating similar wave crest shape.

Initial pulse parameters are: $P_0$ = 15.66 W, $T_{FWHM}$ = 105 fs, $\lambda_1$ = 820 nm, $\lambda_2$ = 755 nm. N.B: The maximum height of the crest, as well as it's lengths along the Z-axis is very sensitive to initial conditions. For example a little larger wavelengths difference would make the crest's peak spike-like typical for soliton's collisions [19, 20], while slightly smaller wavelength separation would make the crest longer lasting over Z, but with height less pronounced. For the example above delicate initial conditions were specially provided to observe a fusion.

In the following we demonstrate how such conditions could be met in a real-life process of SC generation. The process of SC formation provides a brilliant opportunity to study various solitons interactions, such as bound state formation [21, 22], generation of giant dispersive wave during soliton's collision [23]. In order to simulate a process of broadband SC high power $T_{FWHM}$ = 105 fs pulses ranging from 50 kW to 150 kW, corresponding to soliton's order from 56 to 97 were injected with small background noise into the fiber, described above. The initial fission of higher order soliton [24] results in a plethora of quasi-fundamental red-shifted solitons accompanied by strong dispersive waves. Solitons strongly accelerated by radiation emitted from preceding ones may collide, resulting in an effective fusion. Interestingly, the concept of accelerating solitons with bended trajectories was studied in a broader class of physical models [25, 26].

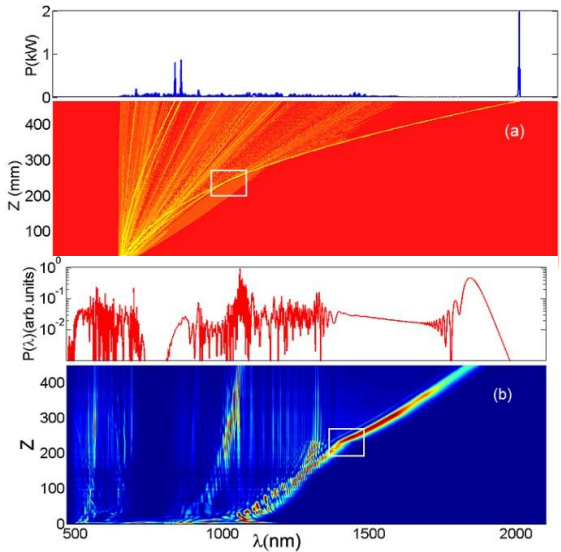

Fig. 2. (Color online) Dynamics of 105-fs, 800-nm pulse with Pin = 100 KW in (a) temporal domain and (b) spectral domains. The region of soliton collision is highlighted.

In [16] it was proved that, it is the radiation emitted from the stronger red-shifted soliton that changes the trajectory of the trailing one and enables the generation of new frequencies. This mechanism is very sensitive to the phase mismatch [11] and may provide gradual enough decrease of temporal shift between two fission ejected solitons until they meet in fusion-like process rather than in regular quasi-elastic collisions [15, 19, 20].

Fig. 2(a) demonstrates a fusion event, when two solitons combine after the propagation distance of about 235 mm into a single robust structure with enhanced energy and acceleration. At the fiber output after the propagation length of 460 mm the fusion product experiences significant temporal shift. It propagates well preserving its shape and intensity of about twice of the parental solitons (upper panel of Fig. 2(a)). In the spectral domain these processes results in development of a new strong band at the long wavelength side (Fig. 2(b)). Detailed characteristics of two collided solitons such as their peak powers and central wavelengths are presented in Fig.3 (a, b) respectively by special filtering of these solitons in temporal and spectral domains. We focus our attention on the region starting from Z=100 mm, when solitons are strongly distinguished up to Z=250 mm, when they propagate in a joint robust structure after the fusion. Up to the distance of Z =160 mm the second soliton gains energy from the first one by absorbing radiated strong dispersive waves [16] and thus experiences strong red-shift from $\lambda_c$ = 1260 nm to $\lambda_c$ = 1400 nm (Fig.3(b)), becoming more red-shifted that the first soliton. From Z = 160 mm to Z= 235 mm, the two solitons continue their motion with weaker energy interchange and red-shifts ratio well described by the well-known formula [15, 27] $\delta\nu_1/\delta\nu_2$ ~ $(<A_1>/<A_2>)^4$ ~ 3. $<A_1>$ and $<A_2>$ are amplitudes of solitons averaged from Z = 160mm to Z= 235 mm. Before solitons collide at Z = 235mm they have power ratio of about 2, wavelengths separation of 110 nm and phase difference measured at pulses peaks of about $\pi/3$. These small differences in pulse parameters provide an effective constructive interference [6] with a resulting merger into a single soliton propagating with enhanced power and red-shift (red curve in Fig.3 (a, b)).

Fig. 3. (Color online) Properties of the first (dashed green curve) and the second (dotted blue curve) red-shifted solitons shown in Fig. 2

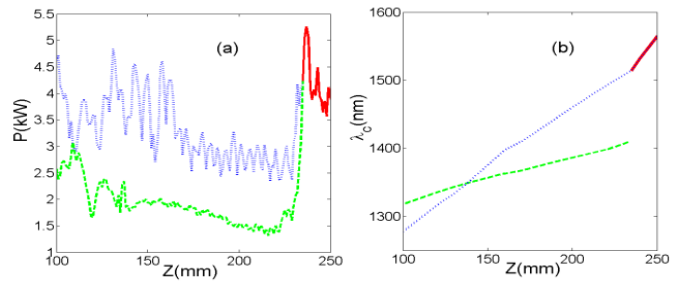

from Z = 100 mm to Z = 250 mm. Displayed: (a) peak powers, (b) central wavelengths. Red curves represent the joint dynamics of the fusion product.

As mentioned above, conditions for the fusion are very delicate and assuming small random noise in input, the event becomes a rare one observed only several times in few hundreds runs with different input noise realizations. It is also worth to point out that fusion events are likely to occur when the light is injected deeper in the anomalous dispersion region ($\lambda_2$ > 790 nm). Soliton fusion product can be enhanced by a sequence of non-elastic solitonic collisions. Fig. 4(a) demonstrates a case of non-perfect fusion occurring around Z = 0.155 m followed by another fusion-like long interaction around Z = 0.165 m. This dynamics was observed when 80-KW, 105-fs pulse was injected in the PCF at 820 nm, with $\beta_2$ more than 3 times higher than in case displayed in Fig.2. Consequently peak powers of fundamental quasi-solitons participating in fusion are several times higher. Snapshots of wave dynamics of Fig. 4(a) are presented in Fig. 4(b) at several illustrative locations. Dashed green curve demonstrates the wave intensity in temporal domain at Z=0.14 m, when three single solitons are still separated before the first interaction. Peak powers of all these three solitons located near (T = 9 ps) are smaller than some other less red-shifted solitons (for example at T = 4 ps and T = 5 ps). Once they collide the peak power of interaction product (T = 9.7 ps) becomes higher than any other soliton (blue dotted curve at Z = 0.156 m). The giant peak arises at Z =

0.165 m when a product of two first solitons collides with the next coming one (the black solid curve).

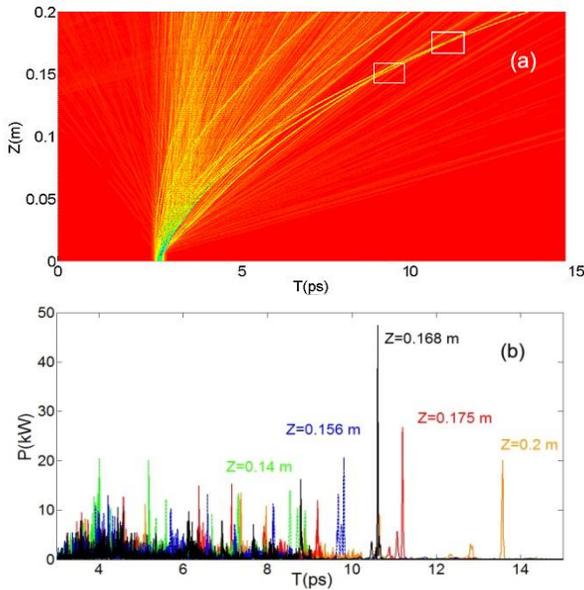

Fig. 4. (Color online)(a) Dynamics of several 105-fs pulses in the PCF in temporal domain with Pin = 80 kW, $\lambda_0$ = 820 nm. (b) Snapshots of wave dynamics at 5 different locations Z. Wave intensities at each of 5 Z-locations are represented with different colors.

With its peak power of about 47 kW it strongly dominates over other solitons. Despite this fusion is also imperfect and followed by some energy loss, the final product dominance is further well pronounced. Examples of snapshots at Z = 0.175 m (solid red curve) and at Z = 0.12 m (solid orange curve) clearly demonstrate the significant intensity difference between the stronger, most temporally delayed fusion product peaks and any other peaks of the same color at the left.

In conclusion, the new candidate for optical rogue waves was presented with pronounced and long lasting domination of wave crest over the background. These rogue waves were created as a result of soliton fusion at the advanced stage of the supercontinuum generation in photonic crystal fiber. It was demonstrated, that acceleration of trailing soliton by the dispersive waves radiated from preceding one may cause to an attraction between the solitons followed by the subsequent fusion. The soliton fusion process requires very delicate conditions for colliding solitons, in particular, small temporal, frequency and phase separation. Thus, small perturbations can easily destroy very sensitive conditions for soliton fusion. Effective fusion and appearance of robust giant rogue wave was observed only several times in hundreds realizations of small noise in the input, which makes such an event rare indeed. Cascaded fusion-like process further significantly enhances the amplitude and the longevity of the optical rogue wave.

### References

[1] D. R. Solli., C. Ropers, P.Koonath, and B. Jalali, Nature (London) **450**, 1054 (2007).
[2] A. Mussot, A. Kudlinski, M. Kolobov, E. Louvergneaux, M. Douay, and M. Taki, Opt. Exp. **17**, 1502 (2009).
[3] J. M. Dudley, G. Genty, F. Dias, B. Kibler, N. Akhmediev, Optics Express **17**, 21497 (2009)
[4] G. Genty, C.M. De Sterke, O. Bang, F. Dias, N. Akhmediev, and J.M. Dudley, Physics Letters A **374** (7), 989 (2010)
[5] M. Erkintalo, G. Genty and J. M. Dudley Eur. Phys.J. Spec. Top. **185** 135 (2010)
[6] A. Antikainen, M. Erkintalo, J. M. Dudley and G. Genty Nonlinearity **25**, R73 (2012)
[7] N. Akhmediev, J.M. Soto-Crespo, and A. Ankiewicz, Phys. Rev. A **80**, 043818 (2009)
[8] N. Akhmediev, J. M. Soto-Crespo, and A. Ankiewicz, Phys. Lett. A **373**, 2137 (2009).
[9] B. Kibler, J. Fatome, C. Finot, G. Millot, F. Dias, G. Genty, N. Akhmediev, J.M. Dudley, Nature Physics **6**, 790 (2010).
[10] J.M.Dudley, G.Genty, and S.Coen, Rev. Mod. Phys. **78** 1135 (2006)
[11] D.V. Skryabin, and A.V. Gorbach, Rev. Mod. Phys. **82** (2), 1287 (2010).
[12] I. Babushkin, A. Husakou, J. Herrmann, and Yuri S. Kivshar, Opt. Express **15** (19), 11978 (2007).
13] F.M. Mitschke and L. F. and Mollenauer, Opt. Lett. **11** (10), 659 (1986).
[14] N. Akhmediev and V.I. Korneev, Theor. Math. Phys. **69**, 1089 (1986).
[15] G. P. Agrawal, Nonlinear Fiber Optics, 4th ed. (Academic Press, 2007).
[16] R. Driben, F. Mitschke, and N. Zhavoronkov, Opt. Express **18** (25), 25993 (2010).
[17] R. Driben and N. Zhavoronkov, Opt. Express **18 (**16), 16733 (2010).
[18] C. Rotschild, B. Alfassi, M. Segev, and O. Cohen, Nature Physics **2**, 769 (2006).
[19] Y. S. Kivshar and B. A. Malomed, Rev. Mod. Phys. **63**, 211 (1991).
[20] R. Driben, and B. A. Malomed, Optics Comm. **197**, 481(2001)
[21] A. Podlipensky, P. Szarniak, N. Y. Joly, C. G. Poulton, andP. St. J. Russell, Opt. Express **15**, 1653 (2007).
[22] A. Podlipensky, P. Szarniak, N. Y. Joly, P. St. J. Russell, J. Opt. Soc. Am. B **25**, 2049(2008).
[23] M. Erkintalo, G. Genty and J. M. Dudley Opt.Lett. **35** 658, (2010)
[24] J. Herrmann, U. Griebner, N. Zhavoronkov, A. Husakou, D. Nickel, J. C. Knight, W. J. Wadsworth, P. St. J. Russell, and G. Korn, Phys. Rev. Lett. **88**, 173901 (2002).
[25] I. Kaminer, R. Bekenstein, J. Nemirovsky, and M. Segev, Phys. Rev. Lett. **108**, 163901 (2012).
[26] ] F. Baronio, A. Degasperis, M. Conforti, and S. Wabnitz, Phys. Rev. Lett. **109**, 044102 (2012).
[27] J. P. Gordon, Opt. Lett. **11**, 662 (1986).